\documentclass[conference]{IEEEtran}
\IEEEoverridecommandlockouts
\usepackage{amsmath,amsfonts}

\usepackage{algorithm}
\usepackage{algorithmic}
\usepackage{multirow}
\usepackage{listings}%
\usepackage{tabularx}
\usepackage{textcomp}
\usepackage{stfloats}
\usepackage{url}
\usepackage{verbatim}
\usepackage{graphicx}
\usepackage{cite}
\usepackage{color}
\UseRawInputEncoding  
\usepackage{subcaption}
\usepackage{booktabs}
\usepackage{caption}
\usepackage{subcaption}
\usepackage{amssymb}

\begin{document}

\title{FM-RME: Foundation Model Empowered \\ Radio Map Estimation
\thanks{This work was supported in part by the US National Science Foundation grants \#2128596, \#2146497, \#2231209, \#2244219, \#2315596, \#2343619, \#2425811,
\#2416872, and \#2413622.}
}
\author{\IEEEauthorblockN{Dong Yang$^\dagger$, \; Yue Wang$^\dagger$, \; Songyang Zhang$^*$, \; Yingshu Li$^\dagger$, \; Zhipeng Cai$^\dagger$, \;Zhi Tian$^\star$}
\IEEEauthorblockA{$^\dagger$Department of Computer Science, Georgia State University, Atlanta, GA, USA\\$^*$Department of Electrical and Computer Engineering, University of Louisiana at Lafayette, LA, USA\\$^\star$Department of Electrical and Computer Engineering, George Mason University, Fairfax, VA, USA}}

\maketitle

\begin{abstract}
Traditional radio map estimation (RME) techniques 
fail to capture  multi-dimensional and dynamic characteristics of complex spectrum environments. Recent data-driven methods  
achieve accurate RME in spatial domain,
but ignore   physical prior knowledge of radio propagation, limiting  data efficiency especially in multi-dimensional scenarios. 
To overcome such limitations, we propose a new foundation model\textcolor{black}{, characterized by self-supervised pre-training on diverse data for zero-shot generalization,} enabling multi-dimensional 
radio map estimation (FM-RME). Specifically, FM-RME builds an effective synergy of two core components: a geometry-aware feature extraction module that encodes physical propagation symmetries, i.e., translation and rotation invariance,
as inductive bias, and an attention-based neural network that learns long-range correlations across the spatial-temporal-spectral domains. 
A masked self-supervised multi-dimensional pre-training strategy is further developed 
to learn generalizable spectrum representations across diverse wireless environments. 
Once pre-trained, FM-RME supports zero-shot inference for multi-dimensional RME, including spatial, temporal, and spectral estimation, without scenario-specific retraining. Simulation results verify that FM-RME exhibits desired learning performance across diverse datasets and zero-shot generalization capabilities beyond existing RME methods.

\end{abstract}


\section{Introduction}
Real-time, high-accuracy spectrum awareness is a fundamental enabler for next-generation wireless systems, facilitating critical applications from dynamic spectrum access to interference mitigation and resource management~\cite{wang2012collecting, wang2012sparsity, yang2024spectrum}. 
Radio map estimation (RME) has emerged as a key technique to provide the panoramic awareness by estimating and visualizing the multi-dimensional distribution of spectrum power~\cite{10757742, romero2022radio}, 
capturing signal strength variations across space, frequency, and time domains. 
Existing RME techniques fall into three classes: model-based, data-driven, and hybrid methods. 
Model-based RME methods use explicit propagation models or statistical assumptions 
to predict spatial signal behaviors. 
However, these models often fail to accurately represent intricate propagation effects in real-world wireless environments, such as complex signal shadowing. 
Alternatively, data-driven RME approaches, using deep neural networks such as U-nets~\cite{levie2021radiounet} and autoencoders~\cite{teganya2021deep}, can learn complex signal propagation patterns from measurements without knowing explicit physical models,
but highly rely on extensive training data. However, spectrum measurements are always sparse and collected irregularly in practice of wireless applications\cite{10505885}, which then degrades their performance dramatically. To solve this challenge, emerging hybrid designs via physics-inspired learning approaches integrate model-based prior knowledge with data-driven methods for RME given limited training data~\cite{yang2025physics, zhang2024physics,  viet2025spatial}.

Despite such advances, the aforementioned RME methods mainly focus on spatial or spectral-spatial domains, making them applicable only for static RME. 
However, in real-world scenarios 
such as real-time UAV logistics, the spectrum conditions are highly dynamic, because transmitters are mobile and can be frequently repositioned. 
This leads to continuous 
changes of spectrum distributions in complex radio environments~\cite{10509639,Wang2024DSL}, rendering static RME inadequate. 
Furthermore, existing data-driven methods are typically trained for specific environmental scenarios and system parameters. As a result, they require retraining when applied 
to new scenarios or configurations, incurring substantial computational and time costs for model retraining. 
While 
 generative AI~\cite{zhao2025temporal} or tensor completion~\cite{chen2025dynamic} 
aim to address temporal dynamics in RME, but they still suffer 
from finite resolution and the requirement for retraining beyond the original region of interest. 

The need for a unified and generalizable model that eliminates frequent retraining motivates a new paradigm shift toward a large-scale radio foundation model.   
To the best of our knowledge, there is no existing work found on large foundation model for multi-dimensional RME.
Intuitively, by pre-training on vast and diverse spectrum datasets, a large foundation model could be used to learn the universal principles of spectrum variation, enabling 
zero-shot adaptation to new and unseen scenarios. 
However, directly applying existing large foundation models to multi-dimensional RME is not effective for the following couple of reasons. 
First, current foundation models are grid-based, by treating the samples of a radio map as pixels of an image under certain resolution, which assumes all the samples of radio map exactly fall on predefined grids. 
Such an assumption imposes fixed spatial resolution and fails to deal with irregularly sampled measurements.
Second, the attention mechanism, as the core of large foundation model architectures, requires substantial data to infer the underlying physical laws of radio propagation. 
In particular, the standard attention mechanism is geometry-agnostic, which lacks awareness of physical 
symmetries governing radio propagation and thus offers no inductive bias toward translation or rotation invariance. 
As a result, two received signals exhibiting identical propagation patterns may be treated as entirely different merely due to differing coordinates or orientations. 
Consequently, a straightforward but naive application of foundation model for RME relies on extremely large amount of training data, making it not only inefficient but also severely limiting its generalization to new scenarios.

In order to address the above challenges, we propose a new foundation model empowered radio map estimation (FM-RME) framework. 
First, we develop a masked self-supervised multi-dimensional pre-training strategy. This allows FM-RME to support multi-dimensional RME and learn general radio map representations from diverse datasets. 
Then, inspired by the physical principles of Maxwell's equations for radio signal propagation, we propose to embed a module defining the geometric invariances in terms of inductive bias into FM-RME to enhance data efficiency. 
Next, an attention-based autoencoder is introduced, which focuses on modeling the long-range correlations across all dimensions to understand the universal structure of the entire radio environment. 
In addition, once the model pre-trained, FM-RME can be applied directly for inference, such as spatial, temporal, and spectral estimation without model retraining and fine-tunings.
The key contributions of this paper are summarized as follows:
\begin{enumerate}
\item We propose the FM-RME framework, as a novel radio foundation model designed for unified multi-dimensional (spatial-spectral-temporal) RME in dynamic environments. To the best of our knowledge, it is the first foundation model designed for RME, which leverages the physics-inspired prior knowledge of radio propagation to holistically integrate self-supervised pretraining, geometry-aware feature extraction, and attention-based autoencoder. 
\item In masked self-supervised pre-training designs, we develop three masked estimation tasks for spatial, temporal, and spectral domains, respectively. They enable FM-RME to effectively capture the inherent spatial-temporal-spectral correlations of spectrum in multi-dimensional RME.  
\item 
For geometry-aware feature extraction module, we develop  
a physics expert, by performing message passing between radio map measurements.
In this way, this module enables FM-RME to
recognize and encode local physical patterns learned from measurements, which are invariant to their absolute position and orientation. 
\item In the design of attention-based autoencoder, we customize both encoder and decoder modules for multi-dimensional RME, by developing spatial-temporal-spectral positional encoding and decoding methods for effectively capturing cross-domain correlations.
\end{enumerate}

\begin{figure} 
\centering
\includegraphics[width=8.8cm]{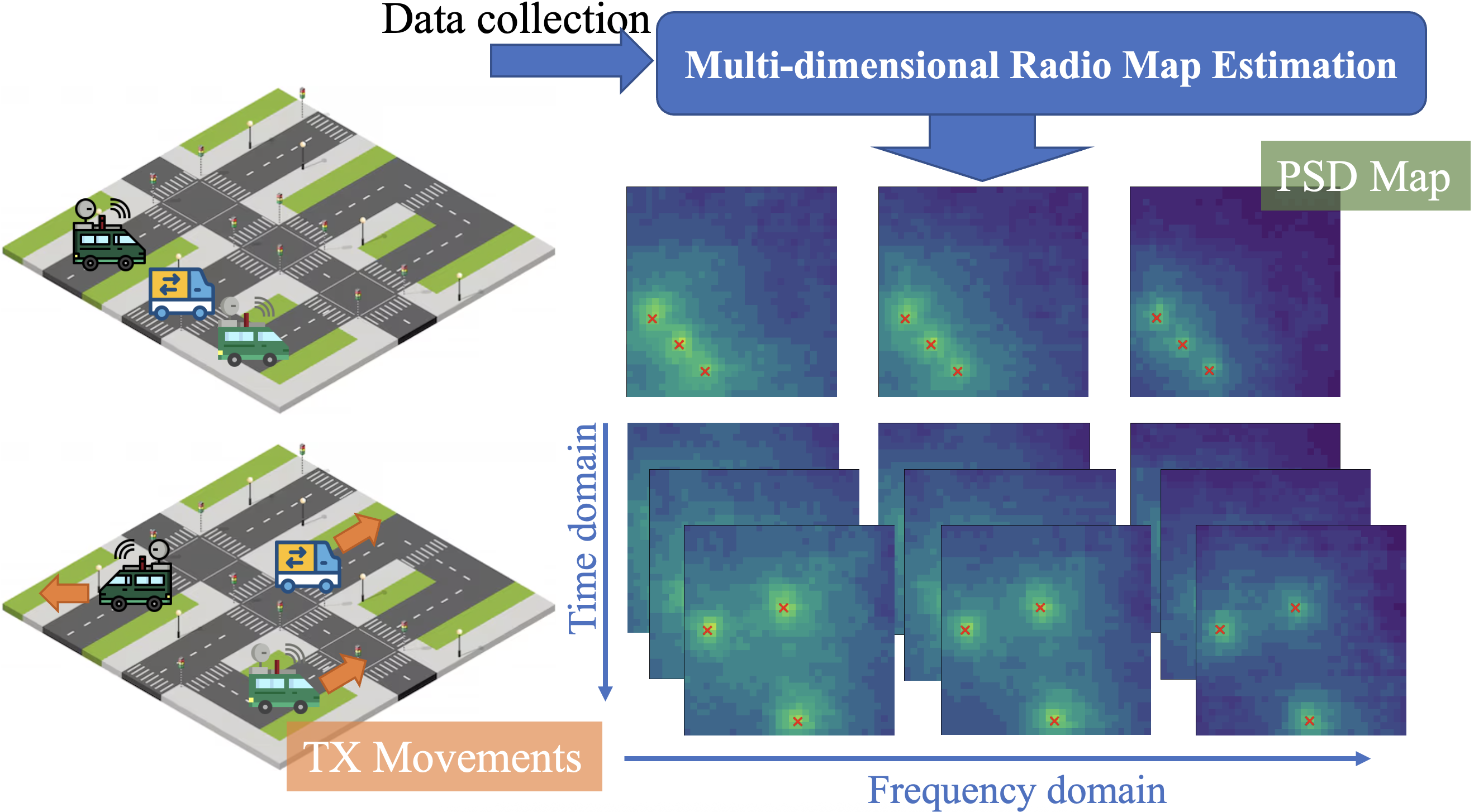}
\caption{Multi-dimensional radio map estimation: RME aims to reconstruct 
full PSD radio maps from sparse measurements, 
enabling RME tasks cross spatial-temporal-spectral domains.
}
\label{fig:Prob_Form}
\end{figure} 

\section{Signal Model and Problem Formulation}
\subsection{Spatial-Temporal-Spectral Radio Map}
We consider practical scenarios in road environments and formulate a physical model to represent the received power spectrum density (PSD) distribution across a target region of interest \( \mathcal{X} \subset \mathbb{R}^2 \), as shown in Fig.~\ref{fig:Prob_Form}. 
To describe the temporal dynamic nature of a mobile network, we first need to model the temporal change of the spectrum over a monitoring period, which is composed of $N_t$ time slots, indexed by $t \in \mathcal{T}=\{t_1,...,t_{N_t}\}$. 
Within each time slot $t$, we adopt a quasi-static assumption, where the number of active transmitters (TXs) and the channel conditions remain constant. 
They can change 
between consecutive time slots, reflecting the mobility of TX and the dynamic nature of the radio environment. 
Let $\mathcal{M}_t$ denote the set of $M_t$ TXs at time slot $t$. The transmitted PSD of the $m$-th TX at frequency $f \in \mathcal{F} = \{ f_1, \dots, f_{N_f} \}$ and within time slot $t$ is denoted by $\Gamma_{m,t}(f)$. The channel frequency response between this TX and a sensor at location \( \boldsymbol{\mathit{x}} \in \mathcal{X} \) and within time slot $t$ is given by \( G_{m,t}(\boldsymbol{\mathit{x}}, f) \). Assuming that the signals from different TXs are not correlated, the received PSD at the location $\boldsymbol{\mathit{x}}$, the frequency $f$, and the time slot $t$ can be expressed as:
\begin{equation}
\label{eq:PSD_y_temporal}
\mathbf{\Phi}(\boldsymbol{\mathit{x}}, f, t) = \sum_{m \in \mathcal{M}_t} \Gamma_{m,t}(f) |G_{m,t}(\boldsymbol{\mathit{x}}, f)|^2 + \eta(\boldsymbol{\mathit{x}}, t, f),
\end{equation}
where $\eta(\boldsymbol{\mathit{x}}, t, f)$ is the PSD of additive white Gaussian noise.

For computational feasibility, we discretize the continuous spatial-spectral-temporal space. The spatial domain $\mathcal{X}$ can be represented by $N_x \times N_y$ locations. Consequently, the ground-truth radio map over the entire monitoring period can be represented as a 4D tensor $\mathbf{\Phi} \in \mathbb{R}^{N_x \times N_y \times N_{t} \times N_f}$, where each voxel $\mathbf{\Phi}(x,y,t,f)$ corresponds to the measured PSD at the $(x, y)$-th spatial location, $t$-th time slot, and $f$-th frequency. 

\subsection{Graph Representation of Radio Map}

To learn physical propagation characteristics that capture inherent geometric correlations in radio maps, we formulate the 4D tensor as a geometric graph $G=(V,E,\mathbf{h}, \mathbf{p})$.
\begin{itemize}
\item Nodes ($V$): Each voxel point $v_i \in V$ in the 4D tensor, indexed by its coordinates $(x_i, y_i, t_i, f_i)$, is treated as a node in the graph.
\item Node scalar feature ($\mathbf{h}$): The initial feature $\mathbf{h}_i^{(0)} \in \mathbb{R}^{3}$ for each node $v_i$ is a scalar representation, containing its measured PSD $\mathbf{\Phi}_i$, its normalized frequency index $f_i^{\text{norm}}$, and its normalized time index $t_i^{\text{norm}}$.
\item Node geometric-vector feature ($\mathbf{p}$): The geometric coordinate of each node $v_i$ is defined by its 2D spatial coordinates $\mathbf{p}_i=[x_i,y_i]\in \mathbb{R}^2$. These vectors are treated as equivariant quantities for the geometry-aware module.
\item Edges ($E$): Graph connectivity is established based on spatial proximity. We use a $k$-nearest neighbors algorithm~\cite{larose2014k} to connect each node to its $k$ closest neighbors in the 2D spatial plane $\mathcal{X}$. Graph structure learning~\cite{yang2024adaptive} and evolutionary algorithms~\cite{liu2023egnn} can also be considered.
\end{itemize}

\subsection{Problem Formulation}
In practical scenarios, we can only acquire a limited subset of the complete 4D radio map tensor $\mathbf{\Phi}$. 
We introduce a binary mask tensor $\mathcal{O} \in \{0,1\}^{N_x \times N_y \times N_t \times N_f}$, where the value is 1 if a measurement is available and 0 otherwise. The available measurements $\tilde{\mathbf{\Phi}}$ can be expressed as:
\begin{equation}
\tilde{\mathbf{\Phi}} = \mathcal{O} \odot \boldsymbol{\Phi},
\end{equation}
where $\odot$ denotes the element-wise product. \textcolor{black}{Then, we adopt a gridless input, treating $\tilde{\mathbf{\Phi}}$ as the set $V_{\text{vis}}=\{\mathbf{h}_i^{(0)}, \mathbf{p}_i\}_{i=1}^{N_{\text{vis}}}$, where $N_{\text{vis}}$ is the number of visible measurements.}

The specific tasks of spatial, temporal, and spectral domain estimation can now be seen as special cases of this general multi-dimensional RME framework, each defined by a different structure of the mask tensor $\mathcal{O}$. Our goal is to develop a 
unified foundation model, represented by a function $f_{\theta}$, that can solve this universal radio map estimation task:
\begin{equation}\label{equ:universal_cartography_task} 
\hat{\boldsymbol{\Phi}} = f_\theta(\textcolor{black}{V_{\text{vis}}}, \mathcal{O}). \end{equation}
This framework unifies multi-dimensional RME:
\begin{itemize}
\item Spatial Estimation: $\mathcal{O}$ contains sparse, random measurements across the spatial domain.
\item Temporal estimation: $\mathcal{O}$ contains sparse measurements for $0< t < T_h$ and no samples for $t \ge T_h$, where $T_h$ is a predefined monitoring time period.
\item Spectral estimation: $\mathcal{O}$ contains sparse measurements randomly collected from some frequencies. 
\end{itemize}
\textcolor{black}{This unified framework allows to pre-train FM-RME for spatial-temporal-spectral RME. And the parameters $\theta$ of FM-RME are optimized via a masked self-supervised objective, minimizing the reconstruction error of masked subsets of $V_{\text{vis}}$.}

\section{FM-RME Architecture}
\begin{figure} 
\centering
\includegraphics[width=8.8cm]{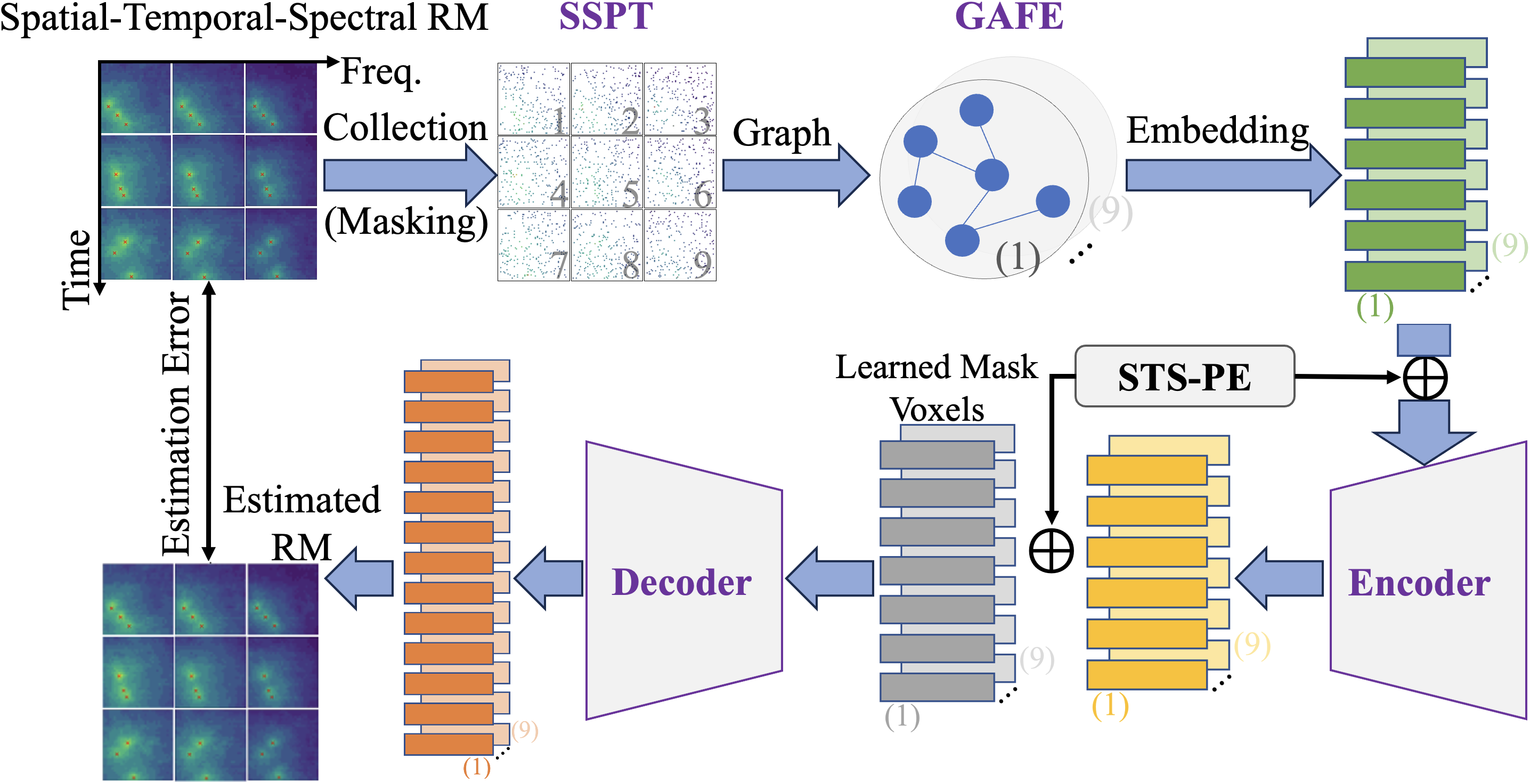}
\caption{The overview of FM-RME framework.
}
\label{fig:framework}
\end{figure}

In this section, a novel radio foundation model FM-RME is introduced to achieve multi-dimensional RME, as shown in Fig.~\ref{fig:framework}. First, the self-supervised pre-training scheme is developed for multi-dimensional radio map estimation task. Then, proposed FM-RME is described in detail, including the geometry-aware feature extraction as a physics expert that encodes physical propagation symmetries, and an attention-based autoencoder as a radio map expert that models across spatial-temporal-spectral correlations and estimates the radio map of the masked location. Next, an inference process for multi-dimensional RME is conducted for  zero-shot generalization.


\subsection{Self-Supervised Pre-Training  (SSPT)}
\textcolor{black}{To create a foundation model capable of generalization,} we adopt a masked self-supervised multi-dimensional pre-training strategy. 
The objective is for the model to learn intrinsic representations of the spectrum structure.

Let all measurements of a full radio map be composed as a set of graph nodes  
$V=\{v_1,v_2,...,v_N\}$ with indices 
$\mathcal{I}= \{1, 2, \ldots, N\}$. During pre-training, we partition the index set $\mathcal{I}$ into: a subset of \textcolor{black}{$N_{\text{vis}}$} visible indices $\mathcal{I}_\text{vis}$ and a subset of \textcolor{black}{$N-N_{\text{vis}}$} masked indices $\mathcal{I}_\text{mask}$, such that $\mathcal{I}=\mathcal{I}_\text{vis} \cup \mathcal{I}_\text{mask}$ and $\mathcal{I}_\text{vis} \cap \mathcal{I}_\text{mask} = \varnothing$. The geometry-aware feature extraction module receives the visible \textcolor{black}{set $V_{\text{vis}}$} as input.

The selection of the masked index set $\mathcal{I}_\text{mask}$ is determined by one of three distinct masking strategies, designed to capture the multi-dimensional variations of the spectrum information:
\subsubsection{Spatial Domain Masked Estimation} To capture spatial correlations, a certain ratio of indices is randomly selected in spatial domain to form $\mathcal{I}_\text{mask}$. This strategy encourages the model to capture a contextual understanding of the information from spectrum in spatial domains.

\subsubsection{Temporal Domain Masked Estimation} To enhance temporal estimation capabilities, indices corresponding to future time steps beyond $T_h$ are assigned to $\mathcal{I}_\text{mask}$. This trains the model to explicitly predict the spectrum evolution based on historical data.

\subsubsection{Spectral Domain Masked Estimation} To learn correlations between frequency bands,  a certain ratio of indices are randomly sampled from some frequencies to construct $\mathcal{I}_\text{mask}$. This is crucial for multi-band estimation task.

\subsection{Geometry-Aware Feature Extraction (GAFE)} 
The principles of Maxwell’s equations dictate that RME exhibits inherent invariance properties, notably with respect to translation and rotation invariance~\cite{viet2025spatial}. 
If the design of a learning-based estimator does not consider and enforce these physical propagation symmetries, such a model has to learn these symmetries from large volume data, which drastically increases the data requirement and also hinders model generalization capability, especially for large foundational models for multi-dimensional RME.

To leverage the physical propagation symmetries, GAFE is designed as a physics expert to encode the geometric symmetries inherent in the radio map into the feature representations. It takes the visible nodes $V_{\text{vis}}$ 
as input and produces a sequence of geometry-aware features. This is achieved by performing $L$ message passing layers, which are designed to be equivariant to spatial rotations and translations.
The update from layer $l$ to $l+1$ is formalized as follows:

\subsubsection{Message Passing}
For each edge $(i,j)\in E$ in $V_\text{vis}$, a message is computed that consists of both a scalar (invariant) part and a vector (equivariant) part.
\paragraph{Scalar Message} An intermediate scalar message $\mathbf{m}_{ij}$ is computed by an multilayer perceptron (MLP) $g_s$. This function aggregates the features of the connected nodes and their squared relative distance:
\begin{equation}\label{equ:scalar_message}
\mathbf{m}_{ij} = g_s(\mathbf{h}^{(l)}_i, \mathbf{h}^{(l)}_j, |\mathbf{p}_i - \mathbf{p}_j|^2, \mathrm{attr}_{ij}),
\end{equation}
where $\mathbf{h}_i, \mathbf{h}_j$ are scalar features, and $\mathrm{attr}_{ij}$ includes invariant features like $|f_i-f_j|$ and $|t_i-t_j|$.
\paragraph{Vector Message} An equivariant vector message $\mathbf{M}_{ij}$ is formed by weighting the relative position vector with a scalar derived from $\mathbf{m}_{ij}$ in \eqref{equ:scalar_message} via the second MLP $g_v$. This operation ensures that the vector's direction is equivariant while its magnitude is modulated by the learned features:
\begin{equation}
\mathbf{M}_{ij} = (\mathbf{p}_i - \mathbf{p}_j) \cdot g_v(\mathbf{m}_{ij}).
\end{equation}
\subsubsection{Aggregation and Node Update}
All messages from neighbors $j\in \mathcal{N}(i)$ at node $i$ are aggregated. The scalar and vector messages are aggregated independently via summation:
\begin{equation}
\textbf{M}_{\text{scalar}, i} = \sum_{j \in \mathcal{N}(i)} \mathbf{m}_{ij},
\end{equation}
\begin{equation}
\textbf{M}_{\text{vector}, i} = \sum_{j \in \mathcal{N}(i)} \mathbf{M}_{ij}.
\end{equation}The scalar features $\mathbf{h}_i$ are updated using a third MLP $g_h$ with a residual connection. The update rule incorporates the aggregated scalar message and the norm of the aggregated vector message $||\mathbf{M}_{\text{vector},i} ||$, which is an invariant scalar summarizing neighborhood geometry:
\begin{equation}\label{eq:node_features}
\textbf{h}^{(l+1)}_i = \textbf{h}^{(l)}_i + g_h ( \text{concat}(\textbf{h}^{(l)}_i, \textbf{M}_{\text{scalar},i}, ||\mathbf{M}_{\text{vector},i} ||) ).
\end{equation}

After $L$ layers, the final node features $\mathbf{h}^{(L)}_i \in \mathbb{R}^{D}$ contain a rich multi-scale summary of the physical propagation symmetries. These feature vectors $\mathbf{H}_\text{vis}=\{\mathbf{h}_i^{(L)} \mid i \in \mathcal{I}_{\text{vis}} \} $ serve as geometry-aware features for the subsequent encoder.

\subsection{Attention-based Autoencoder} 
To effectively capture correlations across spatial, temporal, and spectral domains, we propose an attention-based  encoder-decoder architecture. The encoder processes only the set of visible features calculated by \eqref{eq:node_features}, while a decoder reconstructs the full multi-dimensional radio map voxels.

\subsubsection{Encoder Module}
First, features $\mathbf{H}_\text{vis}$ are augmented with a spatial-temporal-spectral positional encoding (STS-PE), which is conducted via a concatenation of sinusoidal encoding for each feature $\mathbf{h}_i^{(L)}$:
\begin{equation}\label{eq:position_enc_1}
\text{PE}(pos, 2j) = \sin \left( \frac{pos}{10000^{2j/D}} \right),
\end{equation}
\begin{equation}\label{eq:position_enc_2}
\text{PE}(pos, 2j+1) = \cos \left( \frac{pos}{10000^{2j/D}} \right).
\end{equation}
where $pos \in \mathcal{I}_\text{vis}$ is the visible index corresponding to the features, $j = 0, 1, ..., (D/2)-1$ is index of feature dimensions. Therefore, each feature $\mathbf{h}_i^{(L)}$ can calculate a position encoding $\mathbf{P}_{i}$ by \eqref{eq:position_enc_1} and \eqref{eq:position_enc_2}.

The input to the encoder $\mathbf{Z}^{(0)} \in \mathbb{R}^{N_{vis} \times D}$, is formed by adding the position encoding $\mathbf{P}_{i}$ to the visible voxel features $\mathbf{H}_{\text{vis}}$:
$\mathbf{Z}^{(0)}=\mathbf{H}_{\text{vis}}+\mathbf{P}_{i}$. 
Then, the encoder maps the sparse set of visible features $\mathbf{Z}^{(0)}$ to a latent representation. Specifically, the encoder backbone consists of $L_{\text{enc}}$ stacked attention layers. For each layer $l$ from $1$ to $L_{\text{enc}}$, the update is performed using a standard pre-normalization architecture. 
For layer $l$, 
a multi-head self-attention (MHSA) operation is applied to the normalized input, 
followed by a residual addition. 
The intermediate representation $\mathbf{Z}_{\text{int}}^{(l)}$ is calculated as:
\begin{align}\label{eq:Z_int}
\mathbf{Z}_{\text{int}}^{(l)} 
&= \mathbf{Z}^{(l-1)} + \text{MHSA} (\text{LayerNorm} (\mathbf{Z}^{(l-1)})).
\end{align}
Then, a position-wise Feed-Forward Network (FFN) is applied to the normalized intermediate representation, also followed by a residual connection to form the final output of layer~$l$:
\begin{align}\label{eq:Z_l}
\mathbf{Z}^{(l)} 
&= \mathbf{Z}_{\text{int}}^{(l)} + \text{FFN} (\text{LayerNorm} (\mathbf{Z}_{\text{int}}^{(l)} )).
\end{align}
where the FFN is typically designed as a two-layer MLP with activation function $\sigma$:
\begin{align}
    \text{FFN}(\mathbf{X}) = \sigma(\mathbf{X}\mathbf{W}_1 + \mathbf{b}_1)\mathbf{W}_2 + \mathbf{b}_2.
\end{align}
The final output $\mathbf{Z}^{(L_{\text{enc}})}$ of the encoder via \eqref{eq:Z_l} with $l=L_{\text{enc}}$, serves as the latent representation, capturing correlations between visible features across spatial, temporal, and spectral domains.

\subsubsection{Decoder Module}
The decoder module reconstructs the full radio map tensor $\hat{\boldsymbol{\Phi}}$ from $\mathbf{Z}^{(L_{\text{enc}})}$. Specifically, the input to the decoder, $\mathbf{Y}^{(0)} \in \mathbb{R}^{N \times D'}$, is a full-length sequence constructed by placing the encoded $\mathbf{Z}^{(L_{\text{enc}})}$ at their original positions and inserting a shared, learnable masked features embedding $\mathbf{h}_{\text{mask},i}$ at all masked positions $i \in \mathcal{I}_{\text{mask}}$. 

The full sequence $\mathbf{Y}^{(0)}$ is then augmented with the position encoding $\mathbf{P}$ calculated by \eqref{eq:position_enc_1} and \eqref{eq:position_enc_2} for all $N$ positions:
\begin{align}
    \mathbf{Y}^{(0)} \leftarrow \mathbf{Y}^{(0)} + \mathbf{P}.
\end{align}
The decoder backbone, composed of $L_{\text{dec}}$ stacked layers, 
processes $\mathbf{Y}^{(0)}$. 
For each decoder layer $l$ from 1 to $L_{\text{dec}}$, the updating process is similar to the counterpart of encoder as \eqref{eq:Z_int} and \eqref{eq:Z_l}, in which simply replacing $\mathbf{Z}$ by $\mathbf{Y}$. 

After the final decoder layer, a linear projection head $g_{\text{proj}}$ is applied to the output sequence $\mathbf{Y}^{(L_{\text{dec}})}$ to predict the PSD value for each voxel:
\begin{align}
    \hat{\boldsymbol{\Phi}} = g_{\text{proj}}( \mathbf{Y}^{(L_{\text{dec}})} ).
\end{align}

\subsection{Model Inference}
Once the FM-RME is trained via Sections III.A-C,
it is ready to perform a zero-shot inference for multi-dimensional RME as formulated by \eqref{equ:universal_cartography_task} in an unseen scenario. The inference process is a direct application of FM-RME to reconstruct masked radio map information, without requiring any fine-tuning. The input to inference consists of the sparsely collected measurements, 
indicating the visible voxels to FM-RME. 
The inference output provides the estimated PSD values of radio map, which corresponds to the masked voxels to FM-RME. 

\section{Simulation Results} \label{section: simulation}
This section presents simulation results to evaluate the performance of our FM-RME by running on seven datasets compared to the benchmark methods.

\subsection{Simulation Setups}

\begin{table}[]
\centering
\caption{An illustration of the configurations of the simulated 4D spectrum datasets.}
\begin{tabularx}{\linewidth}{X X p{1.3cm} X X X}
\toprule
Dataset  & $N_f$ & $\Delta f$ (MHz) &  $N_t$ & $\Delta t$ (s) & Square \\
\midrule
 D1  & 5 & 900 & 6 & 1 & 64$\times$64 \\
 \hline
D2 & 5 & 900 & 60 & 6 & 64$\times$64 \\
\hline
D3 & 5 & 900 & 40 & 1 & 128$\times$128 \\
\hline
 D4 & 3 & 1800 & 30 & 2 & 128$\times$128 \\
\hline
 D5 & 4 & 1200 & 30 & 2 & 128$\times$128 \\
\hline
D6 & 3 & 1800 & 60 & 3 & 128$\times$128 \\
\hline
D7& 8& 514 &50 & 1& 64$\times$64 \\

\bottomrule
\end{tabularx}
\label{tab:dataset}
\end{table}

\subsubsection{Datasets}
To comprehensively evaluate the estimation performance of FM-RME across various wireless system settings, we generate  7 datasets, designated D1 to D7. These datasets simulate mobile network scenarios by encompassing a wide array of spatial-temporal-spectral spectrum configurations and wireless environments. The simulated setting is designed for mimicking a region of vehicles that drive along the roads, where propagation characteristics are produced by using the Gudmundson shadowing model 
\cite{teganya2021deep}, involving multiple TXs with initial locations randomly located.

The velocity of mobile TXs is constrained to a range of 10 to 15 m$/$s to emulate vehicle movement, and the transmission power is selected between 5 and 11 dBm. From these datasets, the \textcolor{black}{D1-D6} datasets are utilized for the pre-training, while the \textcolor{black}{D7} is reserved for testing the model's generalization performance. 
\textcolor{black}{The first 6 datasets are shuffled into batches with a certain batch size during the pre-training. 
For each batch, the proposed three masking strategies are utilized. The final loss value for gradient descent is the mean loss of the three masking strategies.}
The detailed simulation parameters for each dataset are presented in Table~\ref{tab:dataset}, where the frequencies are sampled from 2.4 to 6 GHz. 
Prior to use, all radio map samples are standardized based on the mean and variance of dataset.

\begin{figure}[]
\centering
\begin{subfigure}[t]{0.46\linewidth}
    \centering
    \includegraphics[width=1\linewidth]{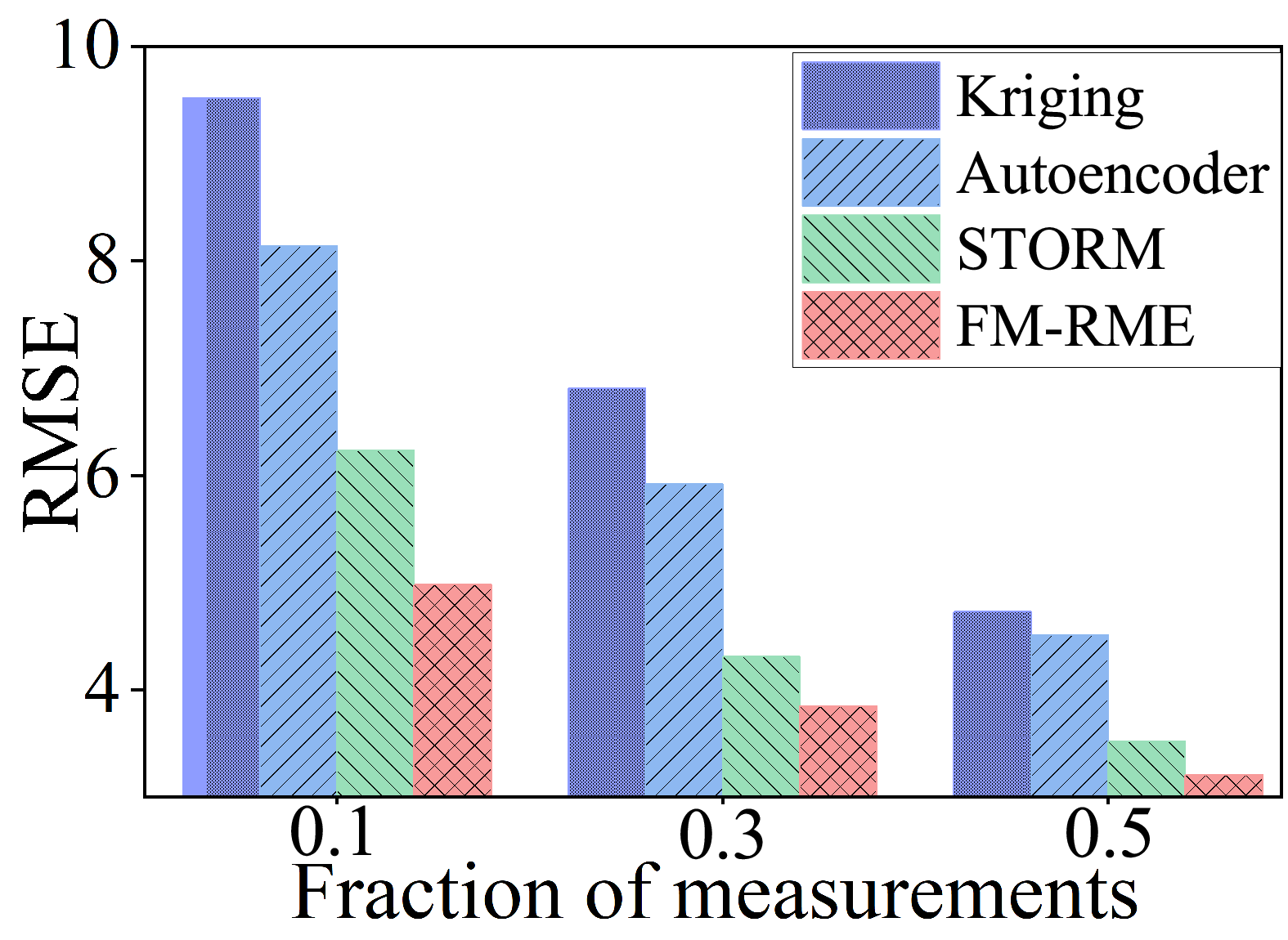}
    \caption{Spatial estimation}
\end{subfigure}
\hfill
\begin{subfigure}[t]{0.473\linewidth}
    \centering
    \includegraphics[width=1\linewidth]{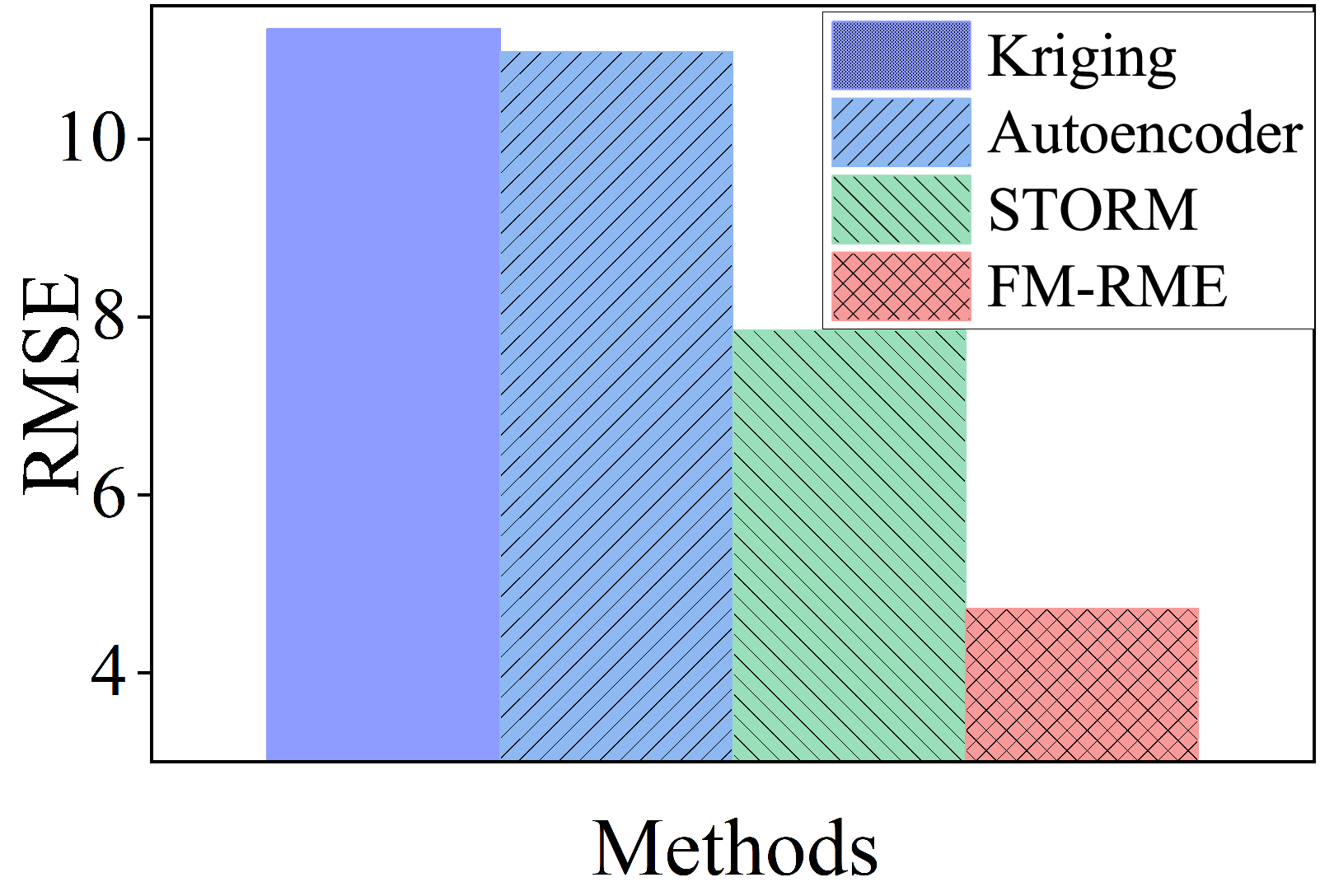}
    \caption{Zero-shot generalization}
\end{subfigure}

\caption{RMSE performance comparison for spatial estimation and zero-shot generalization.} 
\label{fig:spatial_zero_results}
\end{figure}

\subsubsection{Benchmarks}
To evaluate the estimation performance of the proposed FM-RME, we compare it against both model-based and data-driven approaches. Note that baselines are not proposed for the temporal task and that we make some modifications to adapt them accordingly. The baselines are described as follows:
\begin{itemize}
    \item Kriging~\cite{sato2017kriging}: A classical interpolation-based method that models spatial correlation through Gaussian kernels. 
    \item Autoencoder~\cite{teganya2021deep}: A data-driven method with an encoder-decoder architecture is used to estimate radio map.
        \item STORM~\cite{viet2025spatial}: A transformer based model with the translation- and rotation-invariant spatial features is designed for static RME, and ignoring the multi-dimensional correlations.
\end{itemize}

\addtolength{\topmargin}{0.039in}
\subsection{Evaluation and Discussion}
We evaluate the models on four tasks: spatial estimation, zero-shot generalization, temporal, and spectral estimation. The metric for all comparisons is the Root Mean Squared Error (RMSE), which quantifies the difference between the estimated radio map and the ground truth radio map. 
\subsubsection{Spatial Estimation}
For the spatial estimation task, we randomly mask a percentage of voxels from the test samples (from datasets D1-D6) and evaluate the models' reconstruction capability. 
They are tested at 10\%, 30\%, and 50\% measurement sparsity levels, where sparsity indicates the fraction of collected measurements of radio map accessible  to the models. 

Fig~\ref{fig:spatial_zero_results} (a) presents the comparative results for RME on the D1-D6 datasets. Our proposed FM-RME consistently outperforms all benchmarks across all sparsity levels. The performance gap is most significant at the challenging 10\% sparsity level. This demonstrates the significant advantage of proposed foundation model, because FM-RME has learned the inherent geometric correlations in the radio map from the diverse datasets, enabling it to make accurate estimations from very few measurements. 
In contrast, STORM is insufficient for feature representation when dealing with multiple TXs, as it is designed for the single TX scenario. 
The classical Kriging and Autoencoder models both fail to work well, as they lack spatial feature representation from attention-based models.

\subsubsection{Zero-shot Generalization} For the zero-shot generalization task, all models are pre-trained only on datasets D1-D6 and then evaluated directly on the unseen dataset D7, without any fine-tuning, to test their generalization capabilities. Fig~\ref{fig:spatial_zero_results} (b) shows the model's zero-shot generalization to the completely unseen D7 dataset. The D7 dataset features present different temporal and spectral parameters from those in the pre-training datasets. The results show that the benchmark models, having been trained on D1-D6, fail to generalize to new dataset D7, exhibiting extremely high errors. STORM is overfitted to the specific propagation characteristics presented by  the training datasets. 
In contrast, FM-RME shows good generalization performance 
by pre-training on diverse datasets, in which FM-RME has learned the universal principles of spectrum propagation rather than memorizing the overwhelming  characteristics of a specific environment. This allows FM-RME to new scenarios without retraining and finetuning.

\begin{figure}[]
\centering
\begin{subfigure}[t]{0.49\linewidth}
    \centering
    \includegraphics[width=1\linewidth]{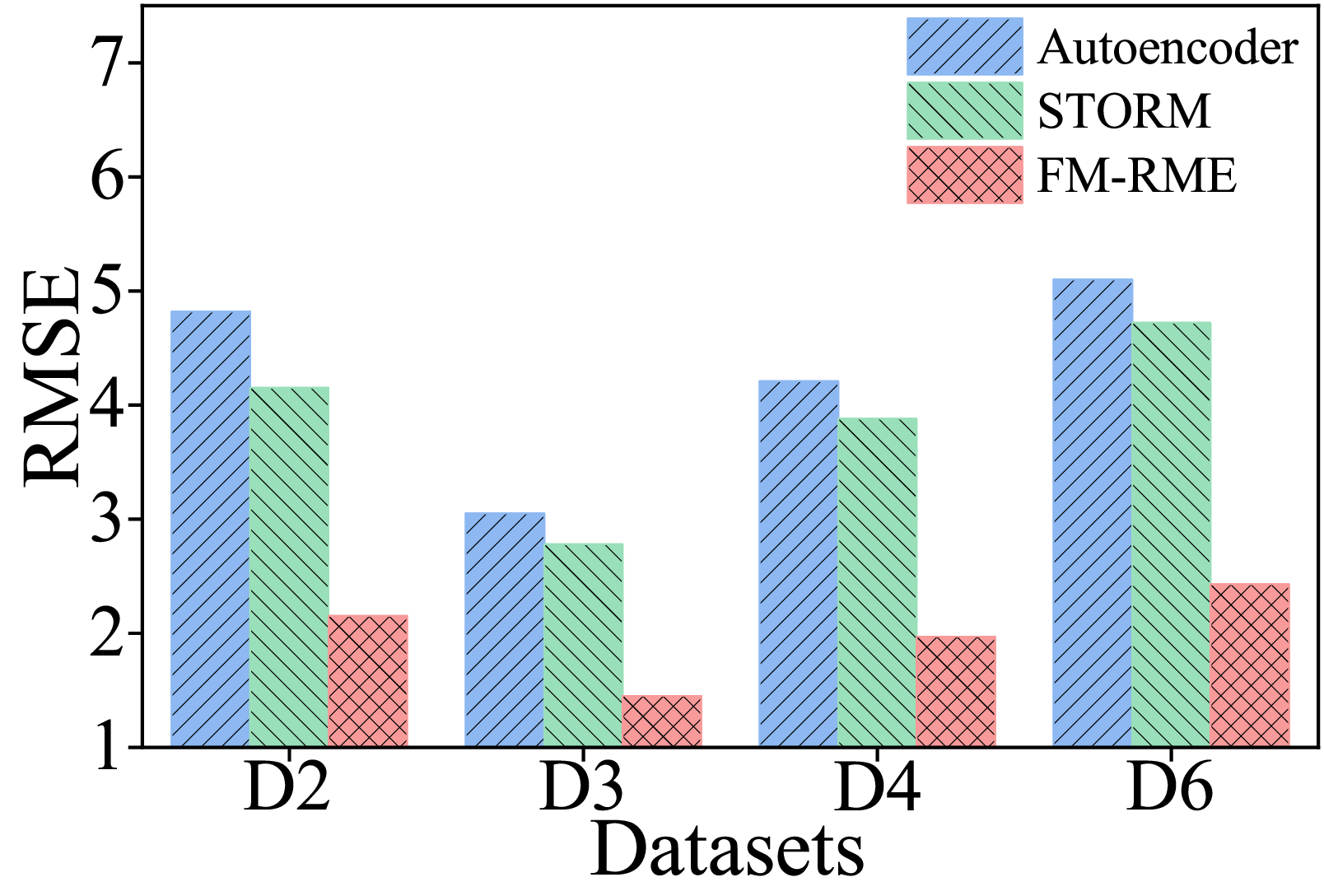}
    \caption{Temporal estimation}
\end{subfigure}
\hfill
\begin{subfigure}[t]{0.49\linewidth}
    \centering
    \includegraphics[width=1\linewidth]{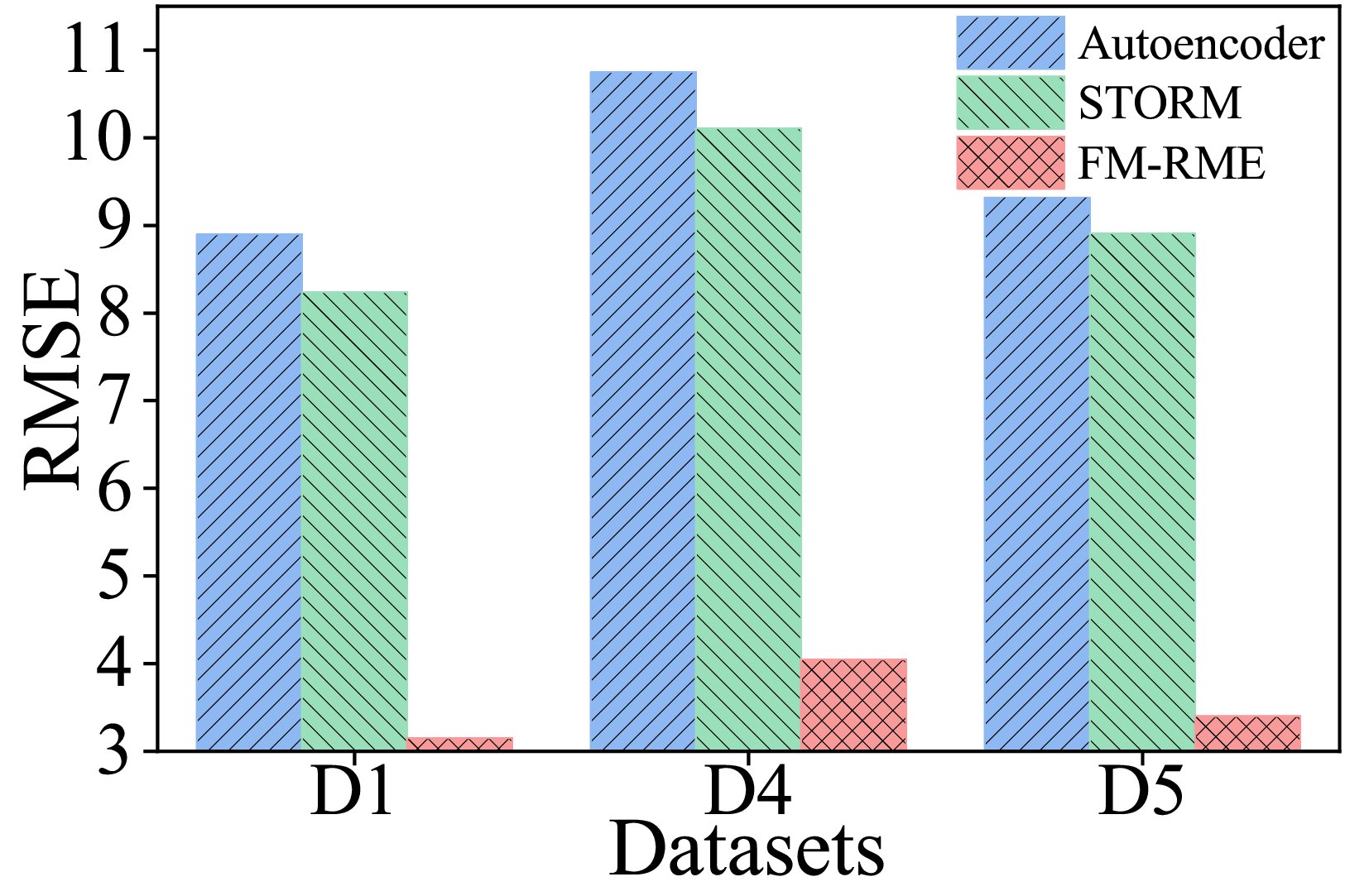}
    \caption{Spectral estimation}
\end{subfigure}
\vspace{-1mm}
\caption{RMSE comparison for temporal and spectral estimation.}
\label{fig:spectral_time_results}
\end{figure}

\subsubsection{Temporal Estimation} A key advantage of FM-RME is its capability to conduct 
temporal estimation which cannot be performed by any other existing benchmark methods. 
To verify that, we train different models on 
D2, D3, D4, D6 datasets having different number of time resolutions. 
Since the Autoencoder and STORM models are not originally designed for temporal estimation, 
they both need to transform the spatial estimation problem over multiple time slots into a sequence-to-sequence manner for temporal RME.  
%
%
%
The results of Fig \ref{fig:spectral_time_results} (a) show that FM-RME outperforms  benchmarks. FM-RME's superior performance stems from its internal representations, allowing to train a more generalizable model than baselines in learning  spectrum fields evolvement over time.

\subsubsection{Spectral Estimation}
We test the different models  to infer across the spectral domain, 
by selecting datasets D1, D4, and D5, which have different frequency bands.
As shown in Fig.~\ref{fig:spectral_time_results} (b), both Autoencoder and STORM fail to capture frequency correlations,  leading to  high estimation errors.
Meanwhile, our FM-RME achieves desired estimation accuracy, by effectively capturing the cross-frequency correlations via well-designed spectral domain masked estimation strategy.


\section{Conclusions}
This paper proposes a foundation model  for multi-dimensional RME in dynamic wireless environments. 
Our FM-RME achieves 
multi-dimensional estimation
by developing a masked self-supervised multi-dimensional pre-training strategy. By representing the radio map as a graph and designing a geometry-aware feature extraction module to capture physical propagation symmetries on the graph, FM-RME recognizes and encodes local
physical patterns learned from measurements. This physics-inspired design, combined with a self-supervised pre-training, geometry-aware feature extraction module, and attention-based autoencoder, allows FM-RME to effectively capture the global spatial-temporal-spectral correlations of spectrum. The simulation results show that FM-RME outperforms the benchmark methods in spatial, temporal, and spectral RME. Further, FM-RME demonstrates superior zero-shot inference capabilities on unseen datasets. It validates that FM-RME learns  fundamental principles of spectrum propagation instead of overfitting to specific environment. 

\bibliographystyle{IEEEtran}
\bibliography{ICC}

\end{document}